\documentclass[preprint,journal]{vgtc}         

\ifpdf
  \pdfoutput=1\relax                   
  \usepackage{graphicx}                
  \DeclareGraphicsExtensions{.pdf,.png,.jpg,.jpeg} 
\else
  \usepackage{graphicx}                
  \DeclareGraphicsExtensions{.eps}     
\fi%

\graphicspath{{figures/}{pictures/}{images/}{./}} 

\usepackage{microtype}                 
\PassOptionsToPackage{warn}{textcomp}  
\usepackage{textcomp}                  
\usepackage{mathptmx}                  
\usepackage{times}                     
\usepackage{cite}                      
\usepackage{tabu}                      
\usepackage{booktabs}                  

\usepackage{subfigure}
\usepackage{xcolor}
\usepackage{bm}
\usepackage{amsmath}
\usepackage{comment}

\definecolor{ruisi}{rgb}{0,0,0}
\definecolor{jingying}{rgb}{0,0,0}
\definecolor{kenny}{rgb}{0,0,0}

\onlineid{1194}

\vgtccategory{Research}
\vgtcpapertype{application/design study}

\title{MienCap: Realtime Performance-Based Facial Animation with Live Mood Dynamics}

\author{
    Ye Pan, Ruisi Zhang, Jingying Wang, Nengfu Chen, Yilin Qiu \\ (Shanghai Jiaotong University, Shanghai, China) \\ 
    Yu Ding (Netease Fuxi AI Lab, Hangzhou, China) \\
    Kenny Mitchell (Edinburgh Napier University, Edinburgh, UK,
    Roblox, CA, USA) \textit{Member, IEEE}
}

\shortauthortitle{Pan \MakeLowercase{\textit{et al.}}: MienCap}

\abstract{Our purpose is to improve performance-based animation which can drive believable 3D stylized characters that are truly perceptual. By combining traditional blendshape animation techniques with multiple machine learning models, we present both non-real time and real time solutions which drive character expressions in a geometrically consistent and perceptually valid way. For the non-real time system, we propose a 3D emotion transfer network makes use of a 2D human image to generate a stylized 3D rig parameters. For the real time system, we propose a blendshape adaption network which generates the character rig parameter motions with geometric consistency and temporally stability. We demonstrate the effectiveness of our system by comparing to a commercial product Faceware. Results reveal that ratings of the recognition, intensity, and attractiveness of expressions depicted for animated characters via our systems are statistically higher than Faceware. Our results may be implemented into the animation pipeline, and provide animators with a system for creating the expressions they wish to use more quickly and accurately.
} 


\CCScatlist{ 
 \CCScat{K.6.1}{Management of Computing and Information Systems}%
{Project and People Management}{Life Cycle};
 \CCScat{K.7.m}{The Computing Profession}{Miscellaneous}{Ethics}
}

\renewcommand{\manuscriptnotetxt}{}

\vgtcinsertpkg

\begin{document}


\firstsection{Introduction}

\maketitle

In computer games and feature films, creating genuine and accurate animated character facial expressions is critical \cite{gonzalez2020rocketbox}. According to the storyline, characters must have readily discernible facial expressions that are congruent with their emotional condition~\cite{porter2000site}. Companies such as Disney and Dreamworks continue to use the classic facial animation workflow, which involves an animator creating expressions at discrete keyframes via an intuitive face rig~\cite{thomas1995illusion,lasseter1987principles}. Character expressions made by animators could be emotive and clear, however, they require significant expertise and a tremendous amount of time.


Motion capture, or “moCap” is an alternative process that provides animators with direct realtime control on 3D characters via performance, providing a more efficient and timely way to generate animation~\cite{li2013realtime,weise2011realtime}. However, the expressive quality and perceptual validity of craft-based animations are frequently lacking in these moCap systems. This is primarily owing to the restricted two assumptions: First, geometric markers are adequate for expression transfer. Second, the geometric markers and features modeled are for human faces rather than stylized character faces. Despite recent breakthroughs in motion capture, modeling capabilities, and control parameterization, existing geometry-based systems fail to address the essential issue of producing clear expressions that can be correctly identified. Relatively few tools or methods~\cite{zhang2020facial,aneja2018learning} exist that assist animators by including the perception of character expressions while creation, however the most majority are not realtime.

We present MienCap, a new method that extends the idea of mocap to the capture of emotional data in real time. We start with proposing a non-realtime (NRT) system, generating stylized character expressions from human performances that is both perceptually and geometrically consistent. We train the 3D expression transfer network that takes images of human faces and generates the rig parameters or controller values of the character that best match the human's facial expression. To increase performance, we develop a two-step filtering strategy to learning the mapping between human and character feature space. We used a lightweight method, called a multi-character adaption network, transfers character expressions to secondary characters.

We then propose a real time (RT) system by combining traditional blendshape animation techniques with a machine learning model. To begin, we train the blendshape adaption network that generates the character rig parameters based on the corresponding blendshape weights. In order to produce temporally stable, flicker-free and geometrically consistent results, we take the rig parameters over the last 3 previous frames, together with the blendshape weights at current frame as the input to our network. The multi-character adaption network is then used to drive secondary characters, allowing us to reuse a principal character rig that we trained in the previous steps.

We investigated the effectiveness of using our methods to animate the characters. We compared three different tracking methods: NRT, RT, and Faceware on the expression recognition, emotion intensity, and the overall attractiveness, as these are crucial factors for audience engagement~\cite{wisessing2020enlighten,hyde2015using}. Results show that both NRT and RT methods significantly improved ratings of the expression recognition, thus validated the effectiveness of 3D expression transfer network and the blendshape adaption network, respectively. 

The following are the key contributions of our work: 
\begin{itemize}
\item For the first time, we contributed a realtime system transferring human facial expressions to multiple 3D stylized characters in a geometrically consistent and perceptually correct way.
\item The amalgamation of data sources (e.g., human expression video database, character blendshape database, controller value database etc.) motivates the further study in the domains of character rigging and animation. In particular, we constructed a high-quality emotional audio-visual dataset as materials for our user study: a set of video clips featuring 2 male and 2 female stylized characters talking with seven basic emotions. 
\item We systematically conducted a user study validated the effectiveness of our solutions in terms of expression recognition, intensity and appeal. This grows the existing knowledge on how animated characters' facial expressions can influence our perception.
\end{itemize}

\section{Related Work}

\subsection{Blendshape facial animation}
An industrial standard for rigging facial animation is the use of blendshapes and may be broadly classed among morphable models \cite{egger2019}. The sum of weighted blendshape models can represent facial expressions quickly and compactly \cite{course}. The neutral phase of an avatar is denoted as $B_0$, a set of its blendshapes are $\{B_1, B_2, ..., B_N\}$ and an avatar with expression can be expressed as eqn.\ref{eqn::blendshape},

\begin{equation}\label{eqn::blendshape}
B = B_0 + \sum_{i=1}^N w_iB_i
\end{equation}
where $w_i$ are blendshape weights. 

Several software tools for markerless face motion capture have already been developed ~\cite{zollhofer2018state,egger2019}. Faceware and Faceshift/ARkit~\cite{arkit}, for instance, collect the blendshapes related to a set of standard expressions given by a human source and map them into stylized characters.

Despite the ease of use of the blendshape representation, there are a few matters to consider. To begin, in order to depict a wide range of emotions, digital artists must frequently construct vast libraries of blendshape targets. For a professional artist, creating a suitably detailed model can take up to a week of labor and numerous cycles of refinement. There have been attempts to automate blendshape construction with individualized comprehensive human facial geometry capture and subsequent optimization procedures~\cite{sumner2004deformation}. However, generating blendshapes for styled avatars still necessitates a time-consuming and labor-intensive modeling process~\cite{onizuka2019landmark,han2021generate}. At the moment, each blendshape is typically hand-crafted by experienced artists using professional applications like Blender or Autodesk MAYA.

Second, in some circumstances, the generally linear structure of blendshapes impacts the quality of the animation~\cite{li2010example,seol2011artist}. Exaggerated expressions or specific expressions would be impossible to portray outside of the linear span~\cite{radzihovsky2020facebaker}. In fact, animators must sometimes account for these shortcomings by sculpting new key shapes or adding new correctives per 3 to 5 frames~\cite{lewis2000pose}.

In light of the success of previous data-driven shape analytical techniques \cite{aneja2018learning,ribera2017facial,chen2020modeling}, we propose a realtime MienCap system that combines blendshape animations with our machine learning techniques to improve expression representation from linear to nonlinear.

\subsection{Data-based animation}
The most related previous work to ours are DeepExpr~\cite{aneja2016modeling}, and ExprGen~\cite{aneja2018learning}.

To begin, DeepExpr~\cite{aneja2016modeling} presented a retrieval approach for locating the closest 2D expression image in the present database to a specific human image, while our NRT MienCap proposed a way for generating for human image a 3D stylized character expression. While DeepExpr inspired us, we employed alternative distance measurements. DeepExpr trained a Convolutional Neural Network (CNN) on a huge human expressions dataset to input a human expression and output the seven classes' probabilities, then trained a similar character model on a character expression dataset, and finally used transfer learning approach to learn a mapping between the human and character feature space. Alternatively, we propose a two-step filtering approach to find the best-matched primary character image with the human face.

Second, our NRT MienCap method shares a similar goal as ExprGen to learn 3D character expressions from individuals in a geometrically consistent and perceptually valid way. ExprGen detects human and character face expressions to provide a perceptual metric for expression generation, and then develops a joint embedding to transfer human expressions to character expressions. We did, however, present a novel method for learning the mapping between the human and character feature space.

Finally and most importantly, we integrate our solution with blendshape animation techniques, to present a realtime (RT) MienCap method. Different from static images, we need to take the time dimension into consideration to avoid flickering output. We further introduced predictive model producing temporal stability and geometric consistency for the character rig parameter motions.

\begin{figure*}[h]
  \centering
  \includegraphics[width=0.8\linewidth]{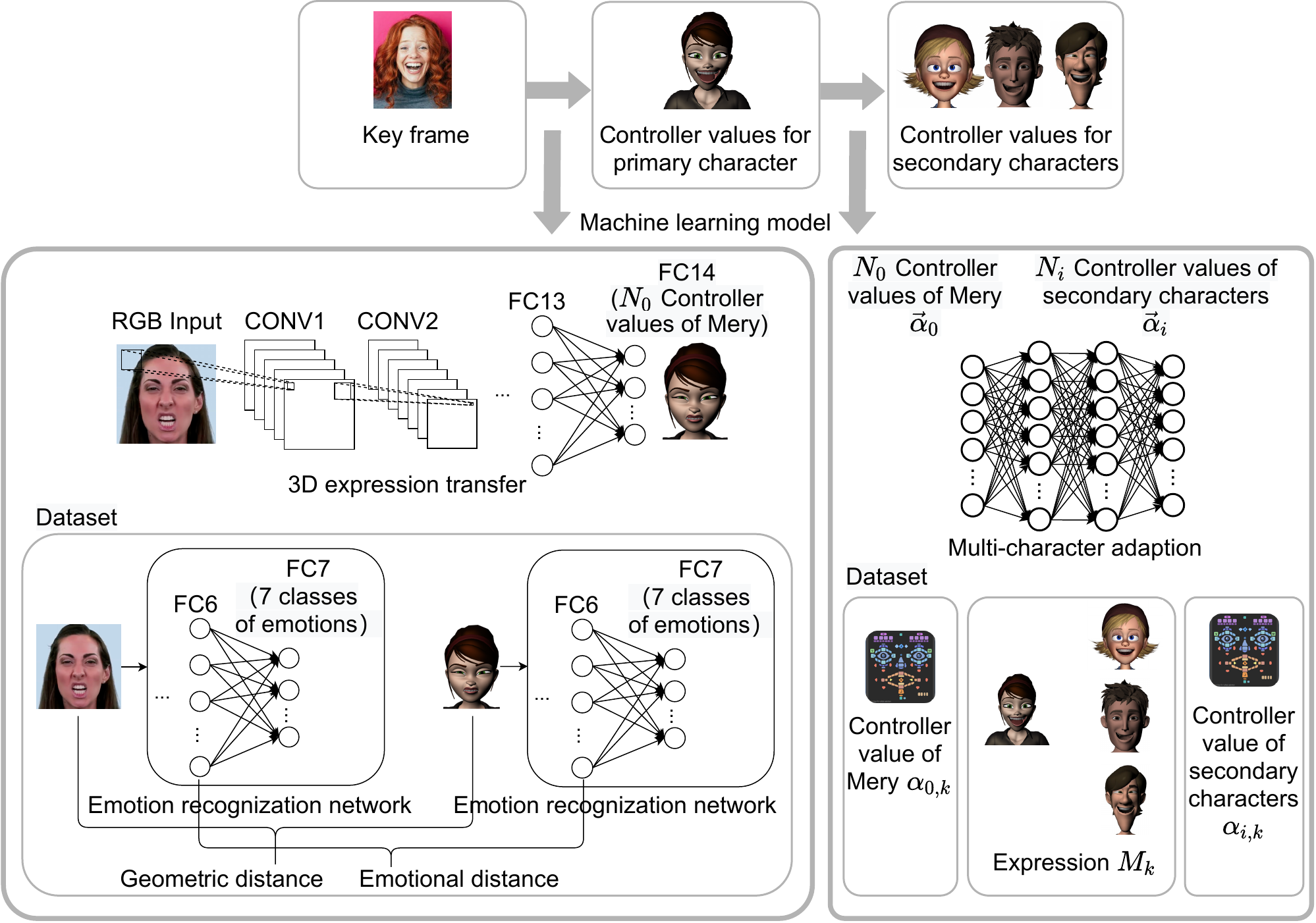}
  \caption{Overview of our NRT MienCap system }
    \label{fig:realTime}
\end{figure*}

\begin{figure*}[h]
  \centering
  \includegraphics[width=0.75\linewidth]{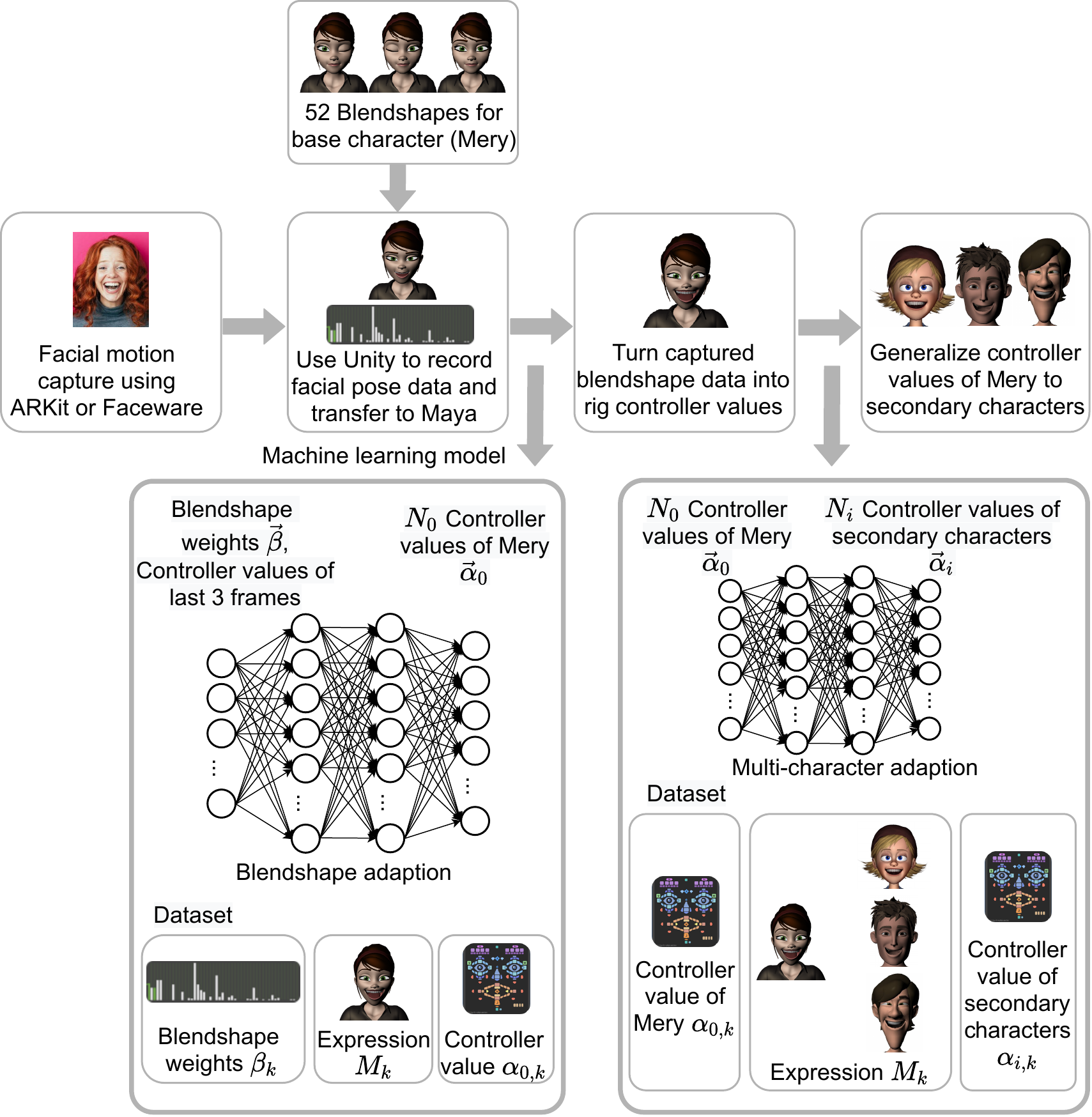}
  \caption{Overview of our RT MienCap system }
    \label{fig:realTime}
\end{figure*}

\section{Non-real time method}
\label{s:NRTSys}
\textcolor{ruisi}{We begin with NRT MienCap system that takes human facial image as input and predicts character's controller values.}
This serves as the building blocks for our RT MienCap system.

\subsection{Data Acquisition}
\label{DataNR}

Our NRT framework employs four databases: (1) Human expression database (HED), (2) Character Expression Database-3D (CED-3D), (3) Character Expression Database-2D (CED-2D), and (4) Human Expression Video Database (HEVD). The specifics of these datasets are as follows:

\textbf{Human Expression Database (HED)}: We created the HED by combining four publicly accessible labeled face expression datasets: (a) the Extended Cohn-Kanade database (CK+)\cite{ck}, (b) the Denver Intensity of Spontaneous Facial Actions (DISFA) database\cite{disfa}, (c) the Karolinska Directed Emotional Faces (KDEF)\cite{kdef}
,  and (d) the MMI database\cite{mmi}. The HED database contains about 100K  images \textcolor{ruisi}{with seven labeled expressions: anger, sadness, joy, neutral, disgust, fear, and surprise}.

\textbf{Character Expression Database-3D (CED-3D)}: We use FERG-3D-DB~\cite{aneja2018learning}, which has \textcolor{ruisi}{about 40000} annotated examples for four stylized characters. 
Each example is a set of controller values that, when applied to the 3D rig, producing a certain facial emotion. Each character's expressions are classified into seven subgroups: anger, sadness, joy, neutral, disgust, fear, and surprise. 

\textbf{Character Expression Database-2D (CED-2D)}: We render the 3D character rigs in the CED-3D into 2D images. \textcolor{ruisi}{When rendering frames, we mark 49 landmark on characters' texture and save the geometric information for each images.} After the characters' faces are cropped and registered with 49 facial landmarks~\cite{xiong2013face}, the images are resized to 256 by 256 pixels for analysis. 

\textbf{Human Expression Video Database (HEVD)}: For training, we make use of the Ryerson Audio-Visual Database of Emotional Speech and Song (RAVDESS) \cite{facedataset}, which includes expressions from 24 professional performers. Anger, sadness, joy, neutral, disgust, fear, and surprise are all expressions. Each expression has two levels of emotional intensity, with an additional neutral expression.

\subsection{Human-Characters Match}

\subsubsection{Human-Primary Character Match}
\label{facePrimary}
\textcolor{ruisi}{To find best-matched primary character image and HED human face pairs}, we proposed a two-step filtering approach: given a input human face query, first retrieve the rendered primary character dataset and find the top 30 character images with the closest \textcolor{ruisi}{emotional distance}; then retrieve matched character image with closest geometric distance among the 30 candidate images.

\textbf{Emotional Distance} 
\textcolor{ruisi}{We use the 512 dimension vector from the fully connected layers of expression classification network proposed in DeepExpr\cite{aneja2016modeling} as Emotion Feature Vector. The expression classification network is first trained on HED and then fine-tuned on CED-2D. For each human-primary character pair, we measure the Jensen—Shannon divergence in Eqn.\ref{math1} as their Emotion Distance.}

\begin{equation}\label{math1}
JSD(H || C)=\frac{1}{2} D(H || M)+\frac{1}{2} D(C || M)
\end{equation}

where $M = \frac{1}{2}(H+C)$, $D(H||M)$ and $D(C||M)$ represents the Kullback—Leibler divergence.

\textbf{Geometric Distance} 
\textcolor{ruisi}{We register 49 facial landmarks from average frontal faces using an affine transformation. Then, we normalize the following geometric distance as Geometric Feature Vector: mouth width (left mouth corner to right mouth corner distance), closed mouth height (distance is vertical between the upper and the lower lip), nose width (distance is horizontal between leftmost and rightmost nose landmarks), left/right eyebrow height (distance is vertical between top of the eyebrow and center of the eye), left/right eyelid height (distance is vertical between top of an eye and bottom of the eye), and left/right lip height (distance is vertical between the lip corner from the lower eyelid). For each human-primary character pair, the L2 norm distance between their  Geometric Feature Vectors are used as Geometric Distance.}

The highlight of our two-step filtering strategy is to further improve the efficiency and accuracy of the retrieved results in a perceptually valid and geometrically consistent way. For example, some expressions, e.g., sad and disgust, the emotional distance is close; while others, e.g., fear and surprise, the geometric distance is close. Thus, the one-step solution combining the emotional distance and the geometric distance together could result in retrieving images with incorrect emotion or far geometry distance.


\subsubsection{Character-Character Match} 
 
Primary character to secondary character matching pairs are retrieved in similar manner: given a input primary character face query, first retrieve rendered secondary character dataset and find the top 30 character images with the closest \textcolor{ruisi}{emotional} distance; then retrieve the matched character image with closest geometric distance among the 30 candidate images.

\subsection{3D \textcolor{ruisi}{Expression Transfer} Network}

\textcolor{ruisi}{We train a 3D expression transfer network in Tab. \ref{3DCharacterTraining}, which takes human faces in HED as input and generates controller values for primary character. The human-primary character pairs are used as ground truth during training.}
\textcolor{ruisi}{The loss function is formulated} in eqn.~\ref{math2}, where $\bm{\alpha}$ is the controller value generated by 3D expression transfer network, $\bm{\alpha}^{'}$ is the controller value in human-primary character pair and $a_{i}$ and $a^{'}_{i}$ are the i-th item in $\bm{\alpha}$ and $\bm{\alpha}^{'}$.

\begin{equation}\label{math2}
HH\left(\bm{\alpha}, \bm{\alpha}^{'}\right)=-\sum_{i} a_{i}^{\prime} \log \left(\operatorname{softmax}\left(a_{i}\right)\right)
\end{equation}

\begin{table}[htbp]
  \centering
  \renewcommand{\arraystretch}{1.05}
  \caption{3D Expression Transfer Network}
  \label{3DCharacterTraining}
  \begin{tabular}{ccc}
    \toprule
    Layers & Patch Size/Stride & Output \\
    \midrule
    CONV1 & 11$\times$11/1 & 64$\times$256$\times$256 \\
    CONV2 & 1$\times$1/2   & 64$\times$128$\times$128 \\
    CONV3 & 5$\times$5/1   & 64$\times$128$\times$128 \\
    CONV4 & 1$\times$1/2   & 64$\times$64$\times$64 \\
    CONV5 & 5$\times$5/1   & 64$\times$64$\times$64 \\
    CONV6 & 1$\times$1/2   & 64$\times$32$\times$32 \\
    CONV7 & 3$\times$3/1   & 64$\times$32$\times$32 \\
    CONV8 & 1$\times$1/2   & 64$\times$16$\times$16 \\
    CONV9 & 3$\times$3/1   & 64$\times$16$\times$16 \\
    CONV10 & 1$\times$1/2  & 64$\times$8$\times$8 \\
    CONV11 & 3$\times$3/1  & 64$\times$8$\times$8 \\
    Avg. Pooling12 & 8$\times$8/1 & 64$\times$1$\times$1 \\
    FC13   & & 1$\times$80\\
    FC14   & & 1$\times$100\\
    \bottomrule   
  \end{tabular}
\end{table}

\begin{figure*}[p]
  \centering
  \includegraphics[width=0.7\linewidth]{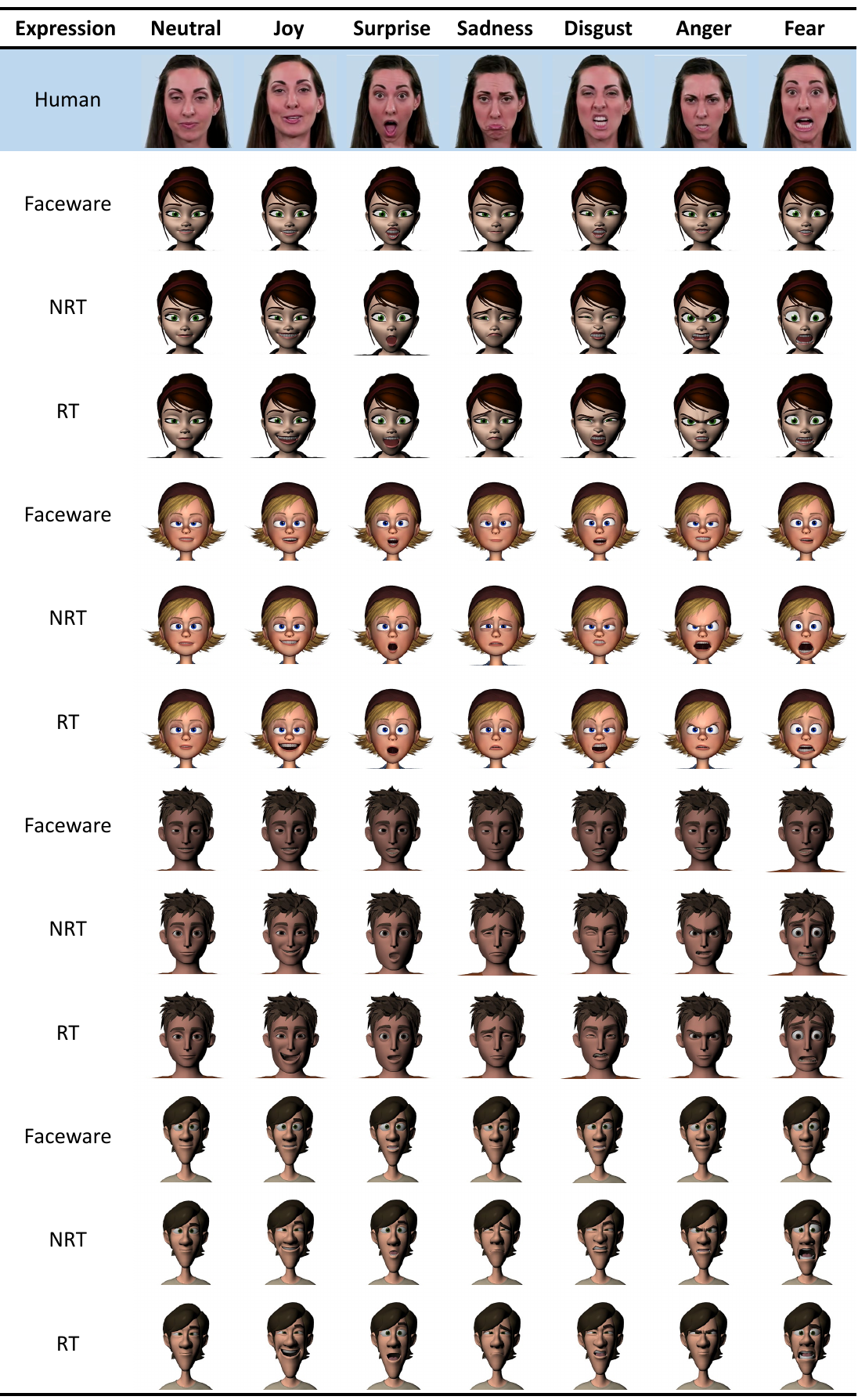}
  \caption{Resulting images generated from Faceware, NRT MienCap, and RT MienCap for all four characters. Uppermost image in each column is input image.}
    \label{fig:realTime}
\end{figure*}

\subsection{\textcolor{ruisi}{Character to Character Transfer}}
Our multi-character adaptation model for NR system is designed to learn to map the controller values of the primary character to the secondary character automatically. We used a lightweight method instead of training a new network for each additional secondary character. We create a separate multilayer perceptron (MLP) for each secondary character, which refers to a two-hidden-layer neural network with N output nodes, M input nodes, and a two-hidden-layer neural network with ReLU activation, where N and M are the number of controller values of the secondary and primary characters, respectively. These networks (together referred to as the Multi-character adaption network) are trained in parallel and then enhanced at the conclusion of the 3D expression transfer network to imprint the input human expression on numerous stylized characters at the same time. 

Note that our multi-character adaptation model is inspired by the C-MLP model of the ExprGen~\cite{aneja2018learning} system. The key difference is methodology of producing a set of matching primary and secondary character pairs (see Sec.3.3.2). Our two-step filtering strategy further ensures the retrieval results that is both perceptually and geometrically consistent. The minor difference is our model is two hidden-layer neural network, whereas C-MLP model is one hidden-layer neural network.

\subsection{Frame Interpolation}
In order to create smooth transitions between images, the convention in animation requires at least 24 frames per second (FPS). However, our NRT \textcolor{ruisi}{MienCap} system only generates at 3 to 4 FPS, since processing each individual frame in deep neural networks takes a considerable amount of time. Thus, the animation would no longer look live or realistic and the user would see images jumping from one expression to another. 

In creating animation clips offline, we simply use linear interpolation to inbetween frames and create more frames to fill the spaces between the original. This allowed to increase this frame rate to 24 FPS without fully solving for the additional frames.

\section{Realtime Method}

The non-realtime method described above is based on static images. It has several drawbacks: (1) The deep neural network leads to a delay when generating key frames. (2) Mapping facial geometry features to emotion space without constraint often produces flickering results.
Therefore, in the realtime method, we designed a light-weight network which takes blendshape weights as input and generates expressive controller values.

Figure \ref{fig:realTime} gives an overview of our pipeline. First, we employed Polywink \cite{polywink} to generate the fullset of blendshapes of our base character rig (Mery). Polywink develops its in-house machine learning technology to automatically generate Facial Action Coding System (FACS) facial rig. Note that we generate the blendshapes only for the sake of Faceware comparison. Our system does not necessarily need to create the fullset of blendshapes to drive the character. Next, we used the award-winning Faceware technology, which is a markless solution for facial motion capture with RGB input. The input image sequence can be either a realtime video stream or a selected video source. The output generated by Faceware Live Client for Unity is a temporal sequence of blendshape weights, which can be directly imported into commercial animation tools. Then, these realtime blendshape weights with a frame rate of 24 fps are fed into the blendshape transfer model to generate controller values for the base character. Lastly, the multiple character adaption model uses the controller values for the base character to produce the controller values for other secondary characters. These output controller values are directly assigned to the controllers of the characters in MAYA.

\subsection{Data Collection}

Our RT framework makes use of two databases: (1) Character Blendshape Database (CBD), and (2) Character Controller Value Database (CCVD).

\textbf{Character Blendshape Database (CBD)}: We collect the blendshape weights of weak emotion intensity videoes in HEVD using Faceware\cite{faceware}. Faceware is a realtime face tracking system, which can effectively caption the geometry feature of human facial poses \textcolor{jingying}{and provide weights of a range of shapes}.  We first calibrate human faces in neutral expression pose and recover the optimized blendshape coefficients (eg. brow down left, brow up left) frame by frame with animation tuning.

\textbf{Character Controller Value Database (CCVD)}: We collect controller value of weak emotion intensity videos in HEVD using the NRT character MienCap system in Section~\ref{s:NRTSys}.

\subsection{Human Expression to Primary Character}\label{sec::human_to_char}

Both the blendshape weights and controller values can animate the characters' expression and some of the parameters are correlated. For example, ``mouthClose" is a coefficient describing closure of the lips independent of jaw position in the blendshape and controller values provide ``up\_lf\_lip\_inout" and ``up\_rf\_lip\_inout" to enable more precise manipulation. However, the frame-by-frame mapping method can cause flickering results. To overcome the inconsistency between frames, \textcolor{ruisi}{we train a blendshape adaption network, which takes blendshape weights and controller value over the last three frames as input and generates controller values in current frame. The controller values in CCVD are used as ground truth during training}. The training process can be formulated as follows:



Given \textcolor{jingying}{an} input vector $\bm{\beta}$ which consists of \textcolor{jingying}{blendshape weights} in \textcolor{jingying}{the} current frame and controller value from \textcolor{jingying}{the} last three frames\textcolor{jingying}{, the} blendshape adaption network $\bm{B}$ \textcolor{jingying}{outputs} the controller value in current frame vector is $\bm{\alpha}^{'} = \bm{B}(\bm{\beta})$.

The loss function of \textcolor{jingying}{the} blendshape adaption network can be formulated as eqn.\ref{que::mlp_definition}, where $\bm{\alpha}$ is the controller value from CCVD, $\bm{\alpha}^{'}$ is the controller value generate\textcolor{jingying}{d} from our blendshape adaption network and, $a_{i}$ and $a^{'}_{i}$ are the i-th item in $\bm{\alpha}$ and $\bm{\alpha}^{'}$.

\begin{equation}\label{que::mlp_definition}
\mathcal{L} ({\bm{\alpha}},{\bm{\alpha}}^{'})=
\sum_{i=0}^{n}\left(a_{i}-a^{'}_{i}\right)^{2}
\end{equation}


\subsection{Character to Character Transfer}
\textcolor{ruisi}{Our Multi-Character Adaptation Model is designed to generate secondary characters' controller value from primary character automatically. We create a lightweight two-hidden-layer neural network takes primary controller values as input and generates secondary controller values with a ReLU activation. When training the model, gradient descent is used with a mini-batch size of 10 and a learning rate of 0.01 to minimize the square loss between the ground truth and output parameters, where the ground truth is obtained from the character-to-character match pair in Sec.3.3.2. The Multi-character adaption models are trained in parallel to generalize human expression from primary character to secondary characters.}

\begin{figure*}[h]
  \begin{center}
       \subfigure[Recognition for video]{\label{f:recognition_video}\includegraphics[height=0.3\linewidth]{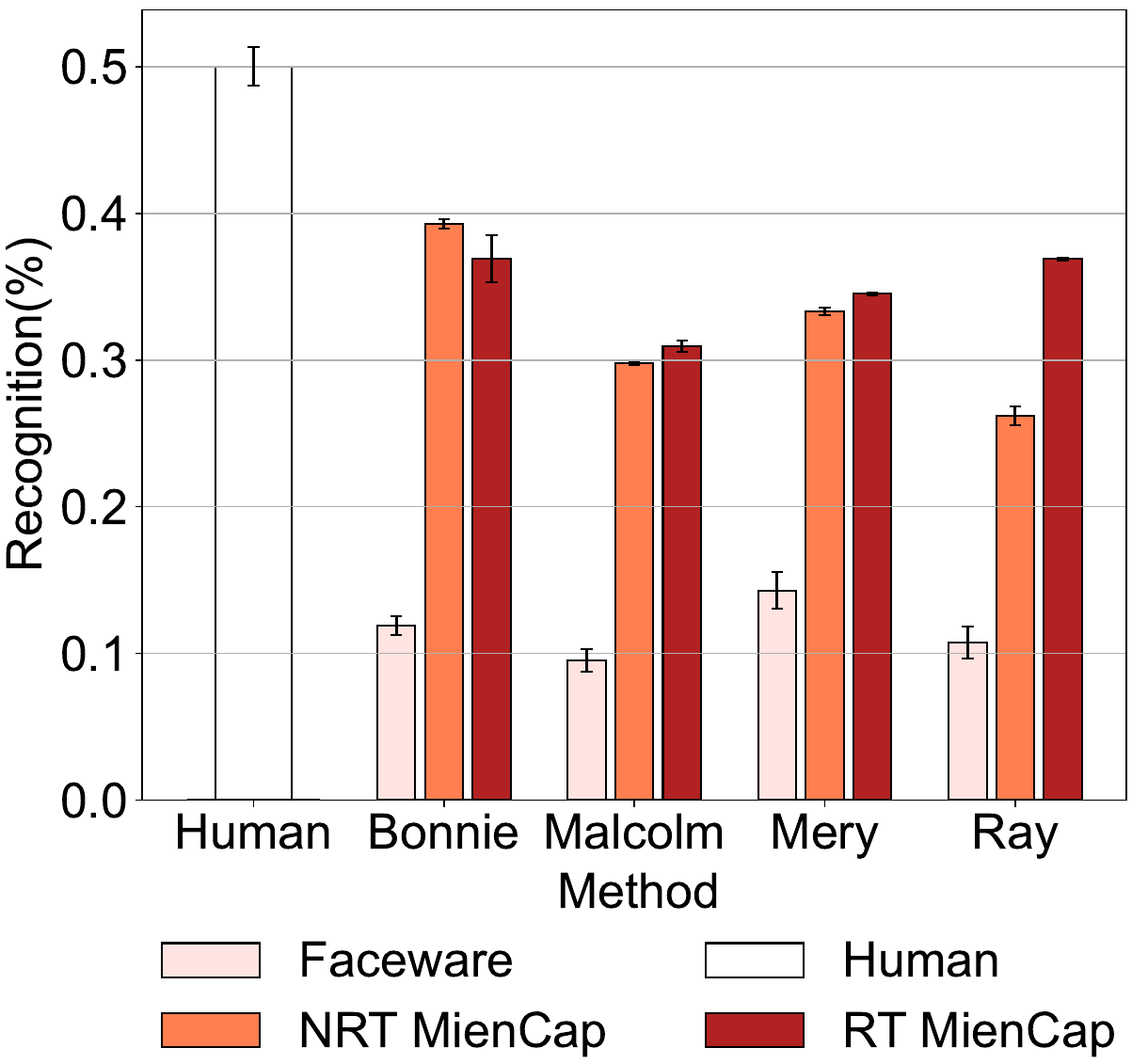}}
    \subfigure[Intensity for video]{\label{f:intensity_video}\includegraphics[height=0.3\linewidth]{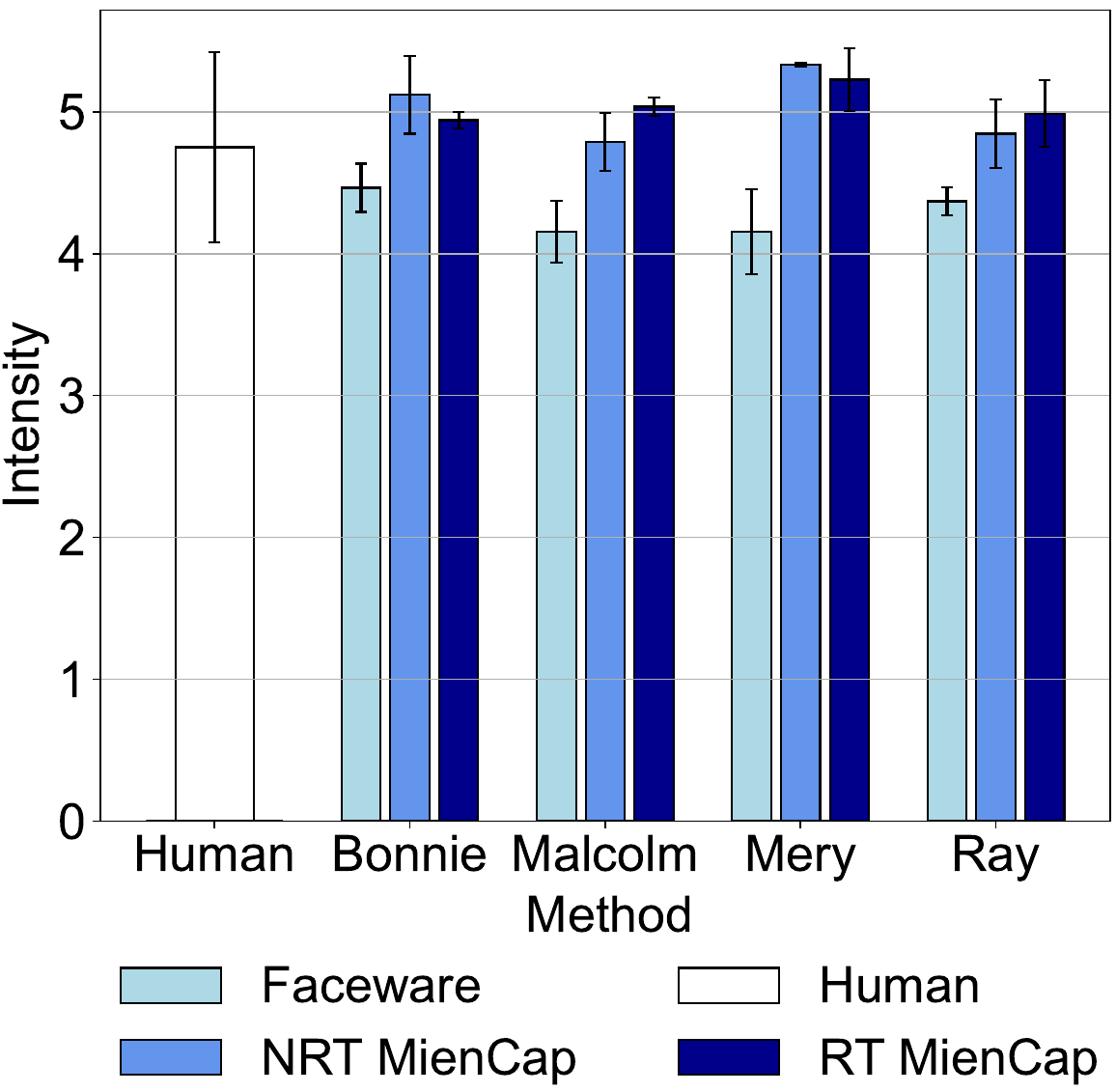}}
  \subfigure[Appeal for video]{\label{f:appeal_video}\includegraphics[height=0.3\linewidth]{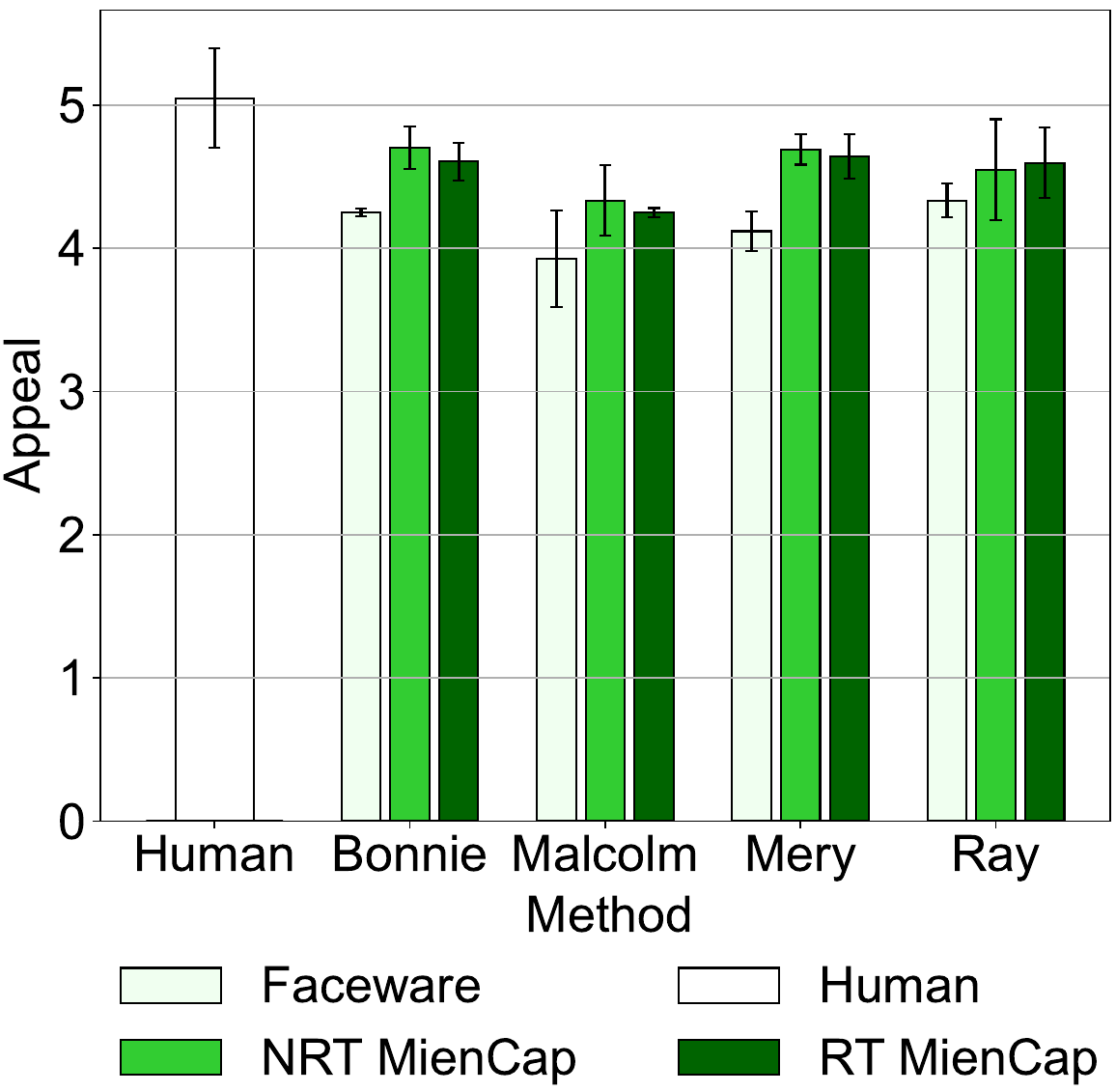}}

  \subfigure[Recognition for image]{\label{f:recognition_image}\includegraphics[height=0.3\linewidth]{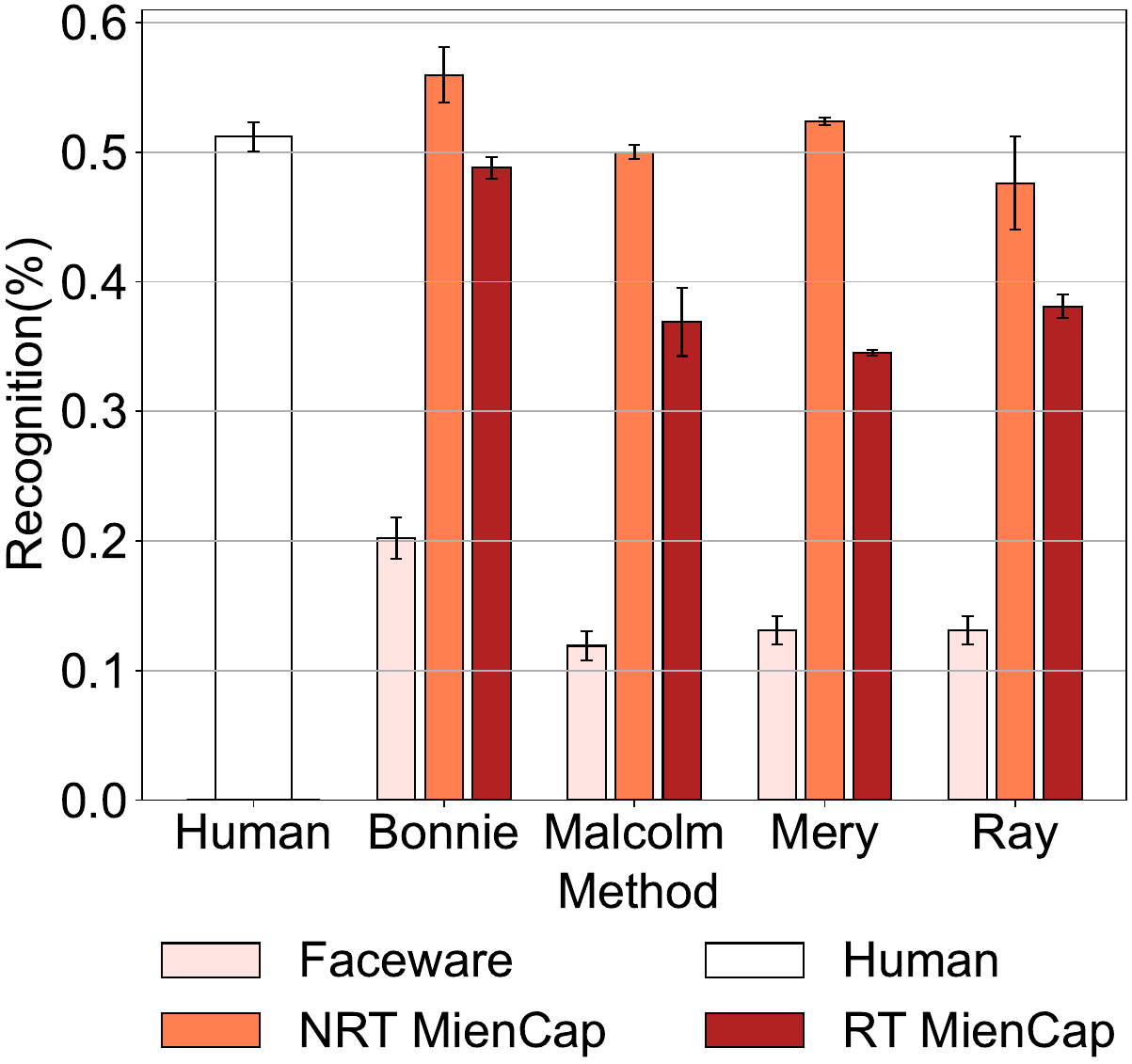}}
    \subfigure[Intensity for image]{\label{f:intensity_image}\includegraphics[height=0.3\linewidth]{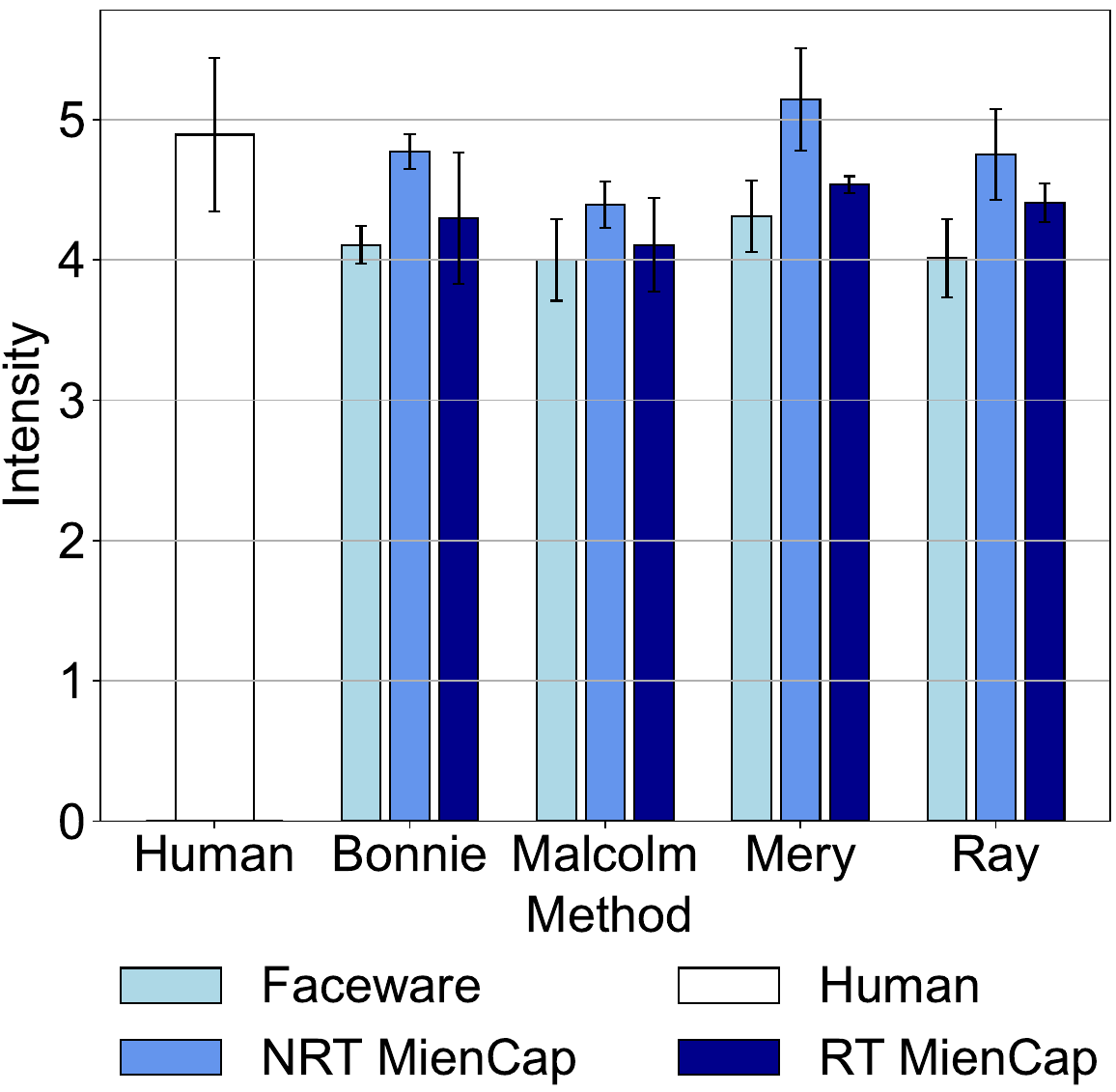}}
    \subfigure[Appeal for image]{\label{f:appeal_image}\includegraphics[height=0.3\linewidth]{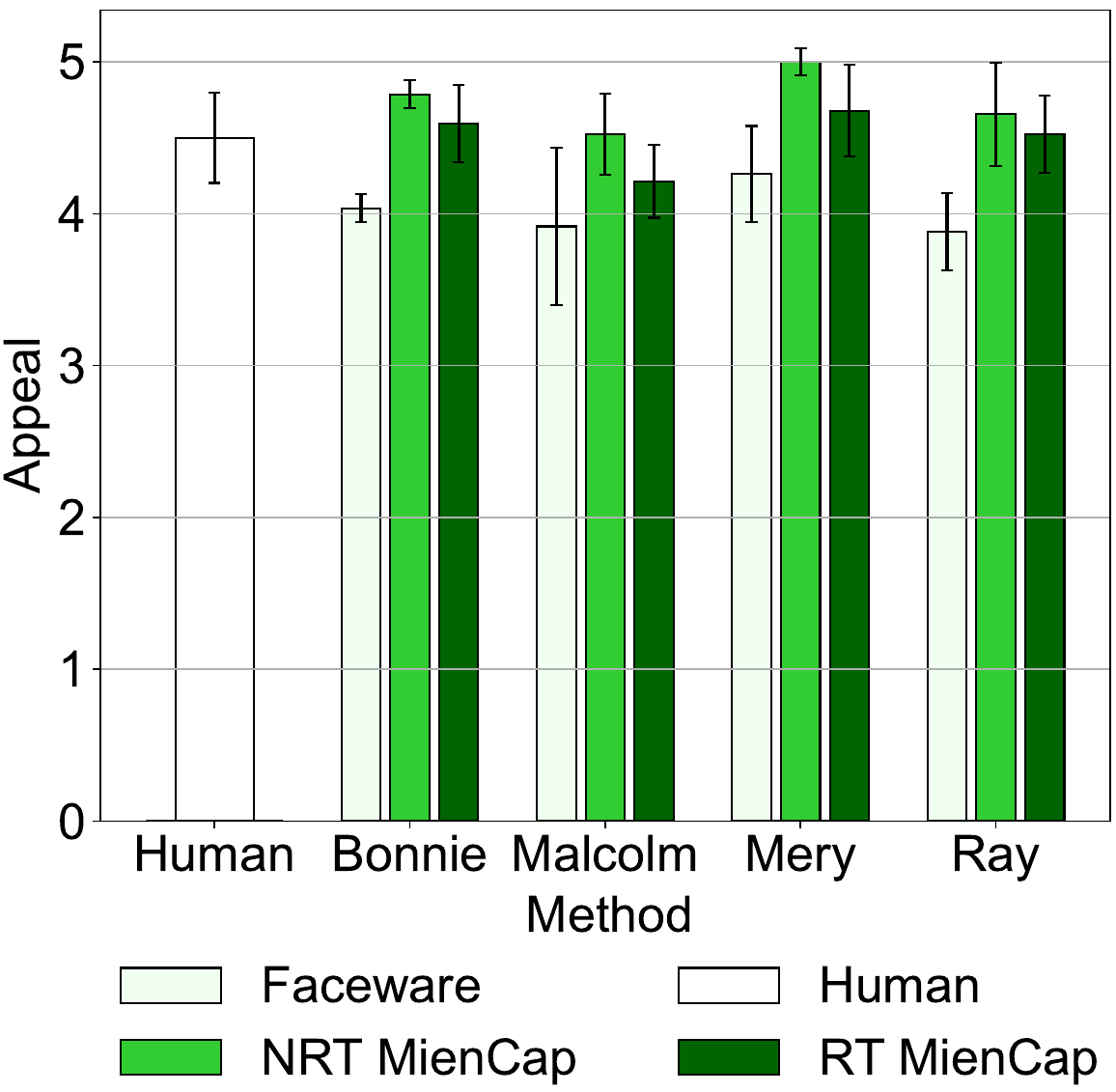}}

  \end{center}
  \caption{Mean for each tracking method and character on recognition, appeal, and intensity.
  Error bars show standard deviation.}
  \label{f:heatMap}
\end{figure*}

\section{Evaluation}
By comparing the expression recognition accuracy of our NRT and RT systems to that of the commercial product Faceware, we were able to assess their effectiveness. Faceware was selected for two purposes: (1) It is the only viable and equivalent system with same input and output modalities as our platforms. Because the results of Faceshift Studio/ARkit require depth sensors to catch human facial motion, we did not compare them. (2) We also used Faceware Live Client for Unity to produce the blendshape weight sequence for our RT MienCap system.

In all the subsequent figures, we use the 2D rendered images to represent the 3D characters, and employ the expression categories in the following manner: N $=$ neutral, A $=$ anger, Sa $=$ sadness, F $=$ fear, D $=$ disgust, J $=$ joy,  Su $=$ surprise.

\subsection{Participant}
We recruited 24 participants from ANON university to completed all conditions of this study. The average age of the participants was 21 years, ranging between 19 and 24 years old; 12 were men. They were naïve to the purposes of the experiment.

\subsection{Material}
\subsubsection{Animation clips}
 We first took alternative 4 set of recordings from 2 male and 2 female actors from HEVD. The dialogue was recorded with 7 basic emotions. The dialogue used for each set of recordings is the same (e.g., “Dogs are sitting by the door”), and we used a face-only format (face, but no voice). Each recording lasted about 3 or 4 seconds. Then, we ran Faceware, NRT \textcolor{ruisi}{MienCap} and RT \textcolor{ruisi}{MienCap} systems to create 4 $\times$ 7 animation clips for 4 characters (Mery\footnote{\textcolor{ruisi}{https://www.meryproject.com}}, Bonnie\footnote{\textcolor{ruisi}{https://www.joshsobelrigs.com/bonnie}}, Ray\footnote{\textcolor{ruisi}{https://www.cgtarian.com}} \& Malcolm\footnote{\textcolor{ruisi}{https://www.animSchool.com}}). 

\subsubsection{Images} 
We used one frame in each recording discussed above and retrieved the same frame via Faceware, NRT \textcolor{ruisi}{MienCap} and RT \textcolor{ruisi}{MienCap} system to create 4 $\times$ 7 images for 4 characters (Mery, Bonnie, Ray \& Malcolm). 

\subsection{Design}
The experiment utilized 4 characters (Mery, Bonnie, Ray \& Malcolm) $\times$ 7 emotions (Neutral, Anger, Sadness, Fear, Disgust, Happiness, \& Surprise) $\times$ 3 capturing methods (Blendshape, Emotion \& Realtime) $\times$ 2 media (Image \& Video) in a mixed design, with a between-subject design for medias, but a within-subject design regarding characters, emotions, and tracking methods. 

Each participant took part in 91 trials to evaluate the input human expression (7 emotions $=$ 7 trials), the generated primary character expression (7 emotions $\times$ 3 capturing methods $=$ 21 trials), and the expression transfer results on different three stylized characters (63 trials). Thus, there were 2184 trials in total. To avoid fatigue or carry-over effects, images or video clips were presented to the participants in random order. 

\subsection{Procedure}
Participants were first presented with an information sheet, and asked to sign a corresponding consent form. They were randomly assigned to either an image condition or a video condition. They were instructed to view an image or animation clip, and then asked to answer three questions:
\begin{itemize}
\item ``Which expression did the character depict?’’ Participants were asked to select one of the words: Neutral, Anger, Sadness, Fear, Disgust, Happiness, Surprise or Other.

\item ``How intense was the indicated emotion depicted by the character?” Participants rated the intensity on a scale from 1 to 7, where 1 represents a rating of ``Not at all'', and 7 represents ``Extremely''. 

\item ``How attractive was the character overall?” Participants rated attractiveness on a scale from 1 to 7, where 1 represents a rating of ``Not at all'', and 7 represents ``Extremely''. 

\end{itemize}

Each participant undertook one practice trial where they could ask questions, and then undertook 91 measured trials.

The participants were paid ANON amount. The experiment took about 30 minutes. The experiment was approved by ANON University Research Ethics Committee.

\begin{figure*}[p]
  \begin{center}
    \subfigure[Human for video]{\label{f:human_video}\includegraphics[width=0.33\linewidth]{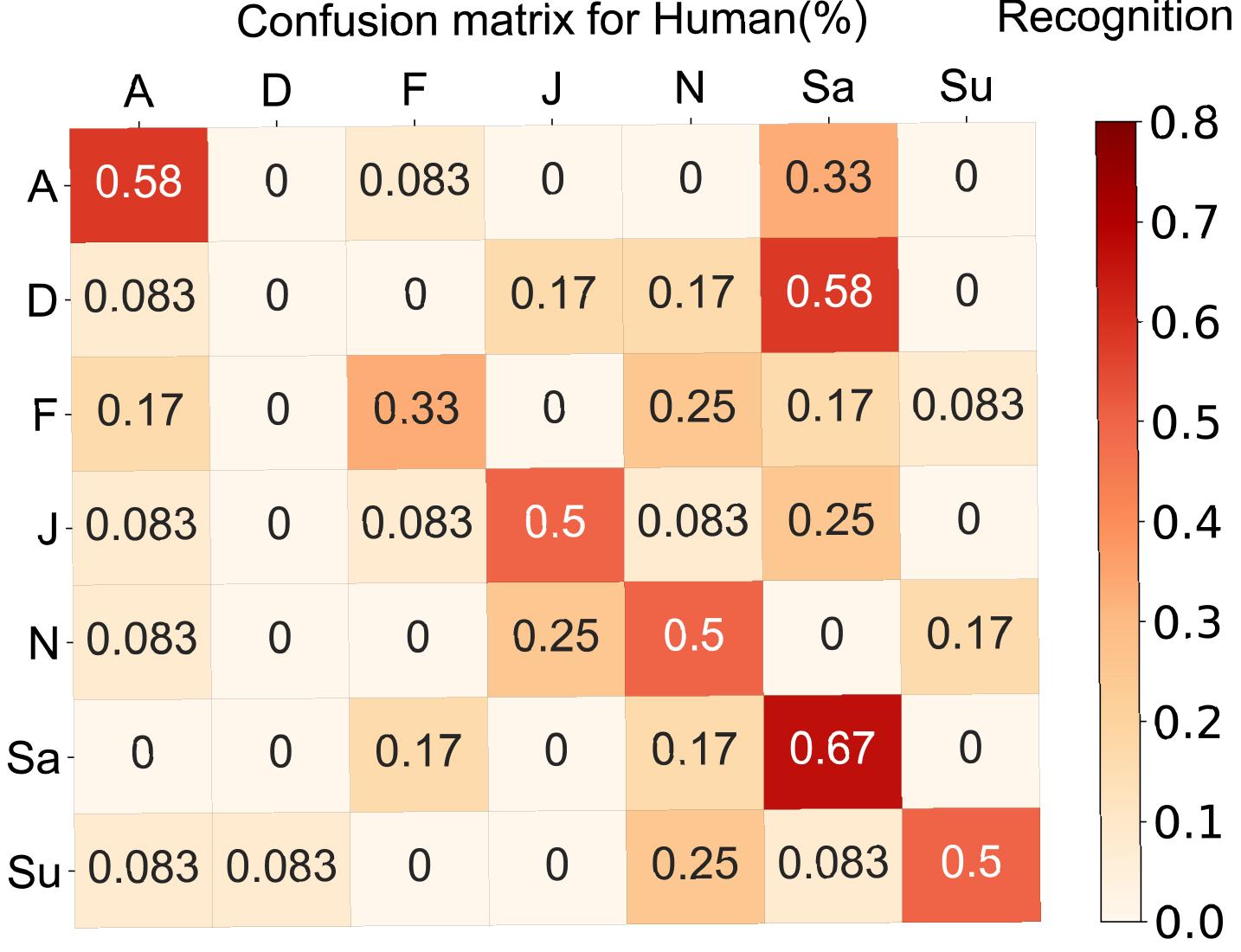}}
    \subfigure[Human for image]{\label{f:human_image}\includegraphics[width=0.33\linewidth]{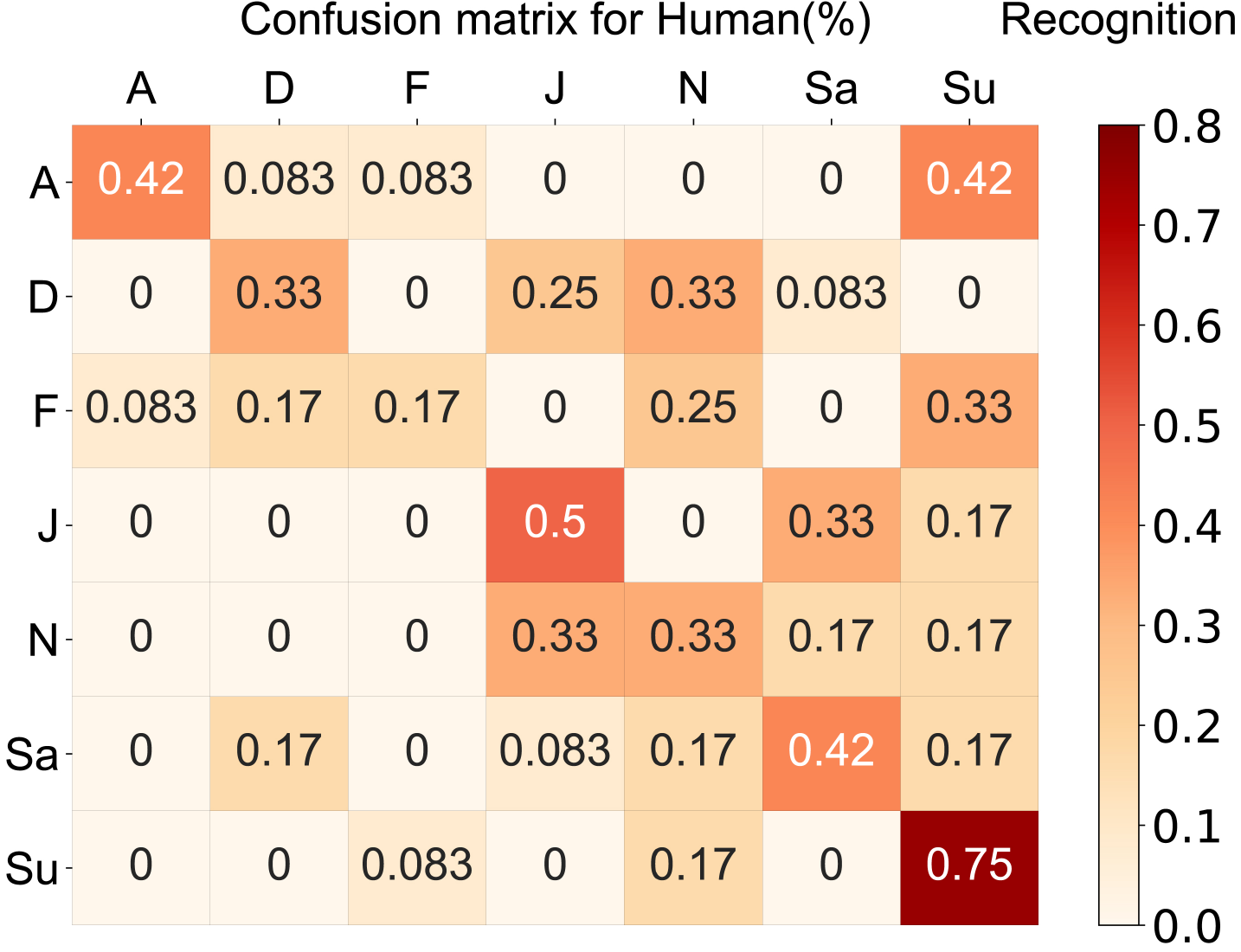}}
    
    \subfigure[Faceware for video]{\label{f:blendshape_video}\includegraphics[width=0.33\linewidth]{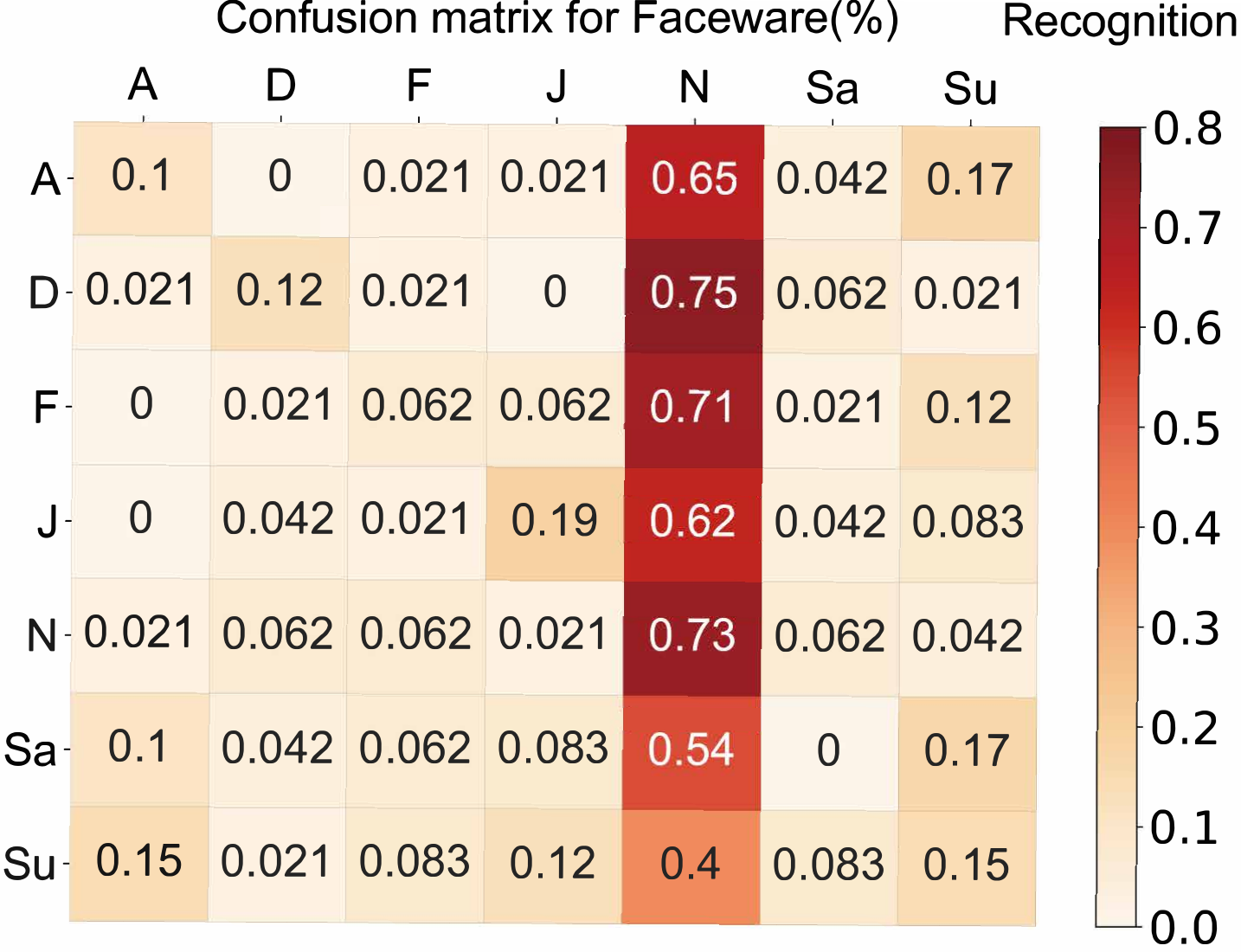}}
    \subfigure[Faceware for image ]{\label{f:blendshape_image}\includegraphics[width=0.33\linewidth]{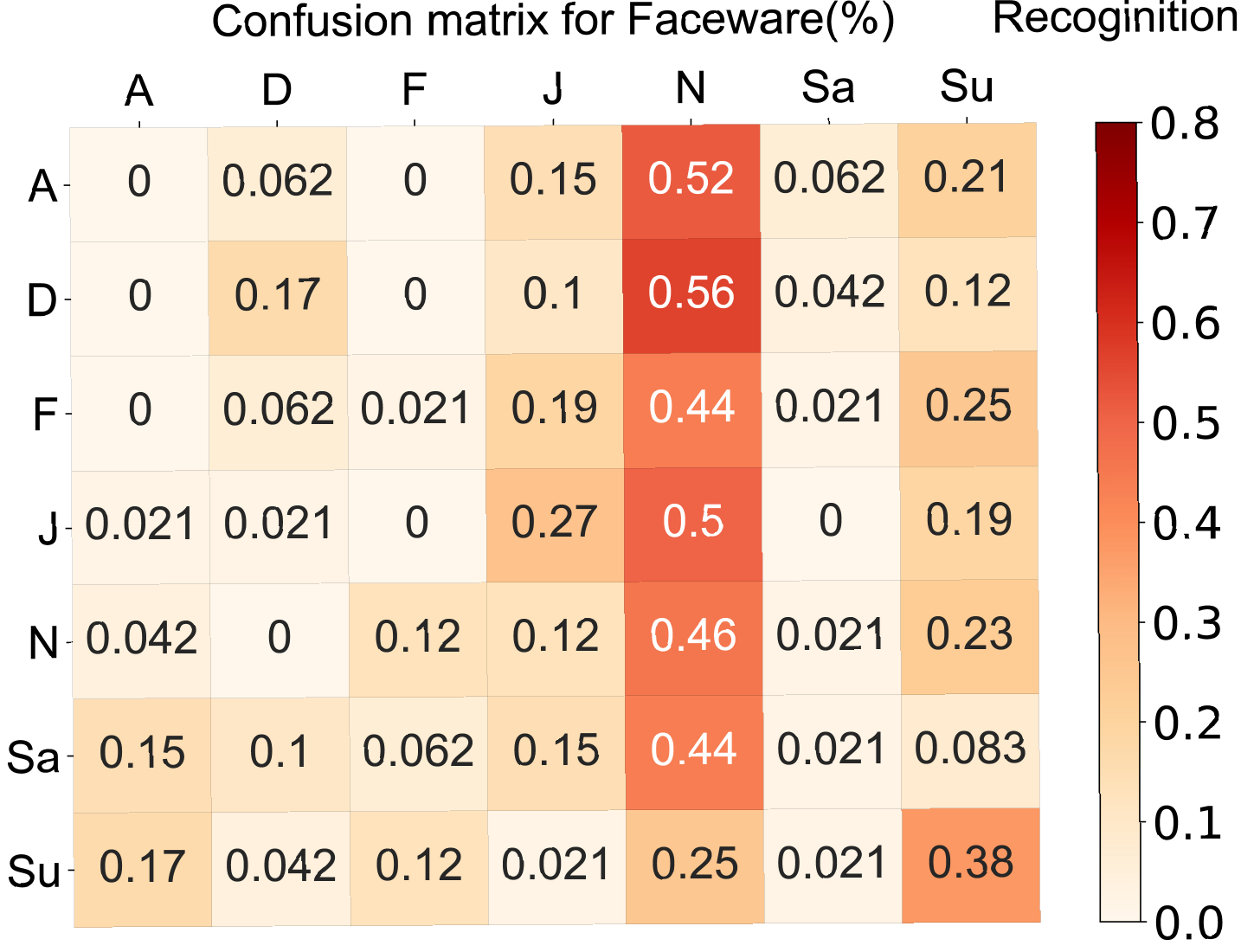}}

  \subfigure[NRT MienCap for video]{\label{f:emotion_video}\includegraphics[width=0.33\linewidth]{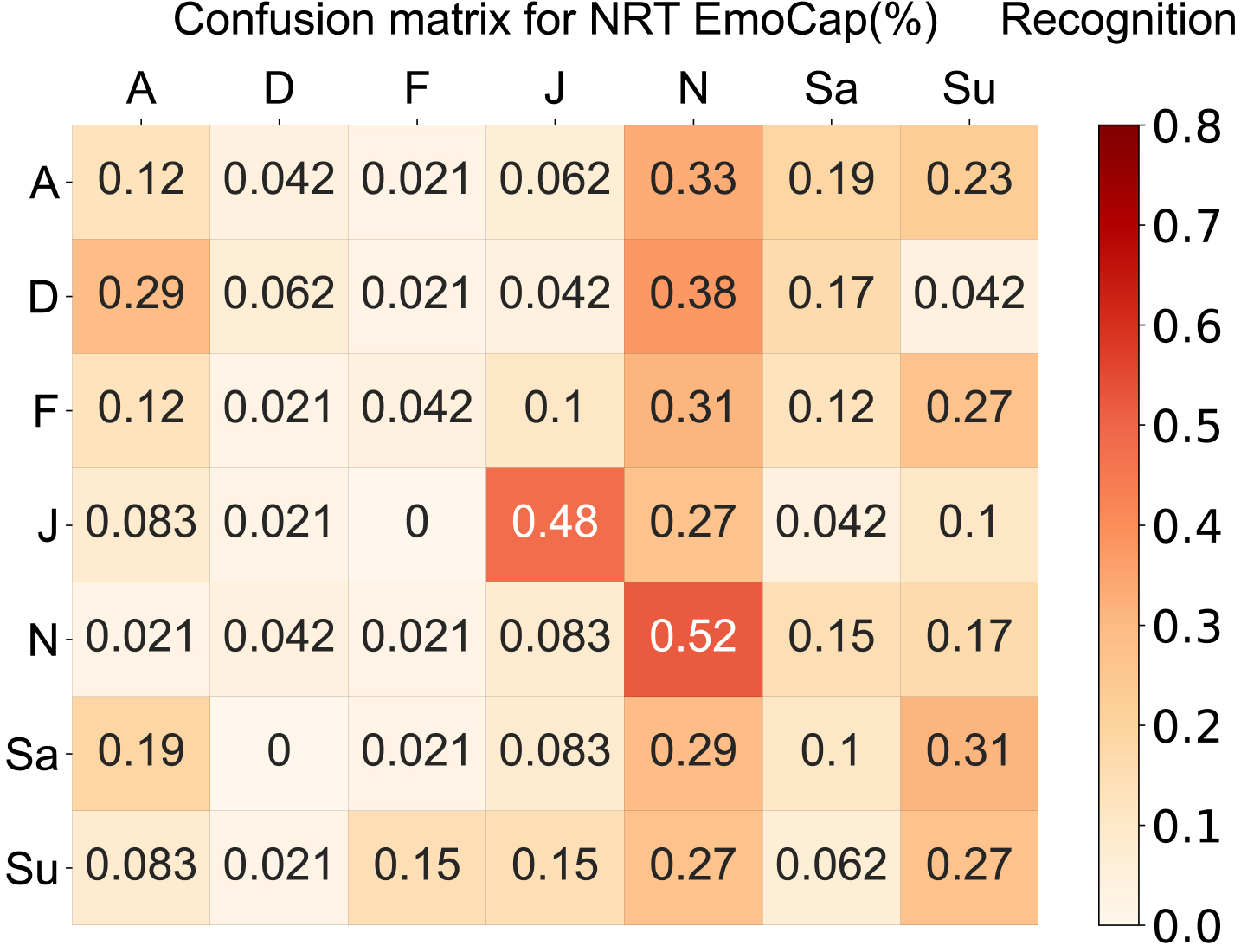}}
    \subfigure[NRT MienCap for image]{\label{f:emotion_image}\includegraphics[width=0.33\linewidth]{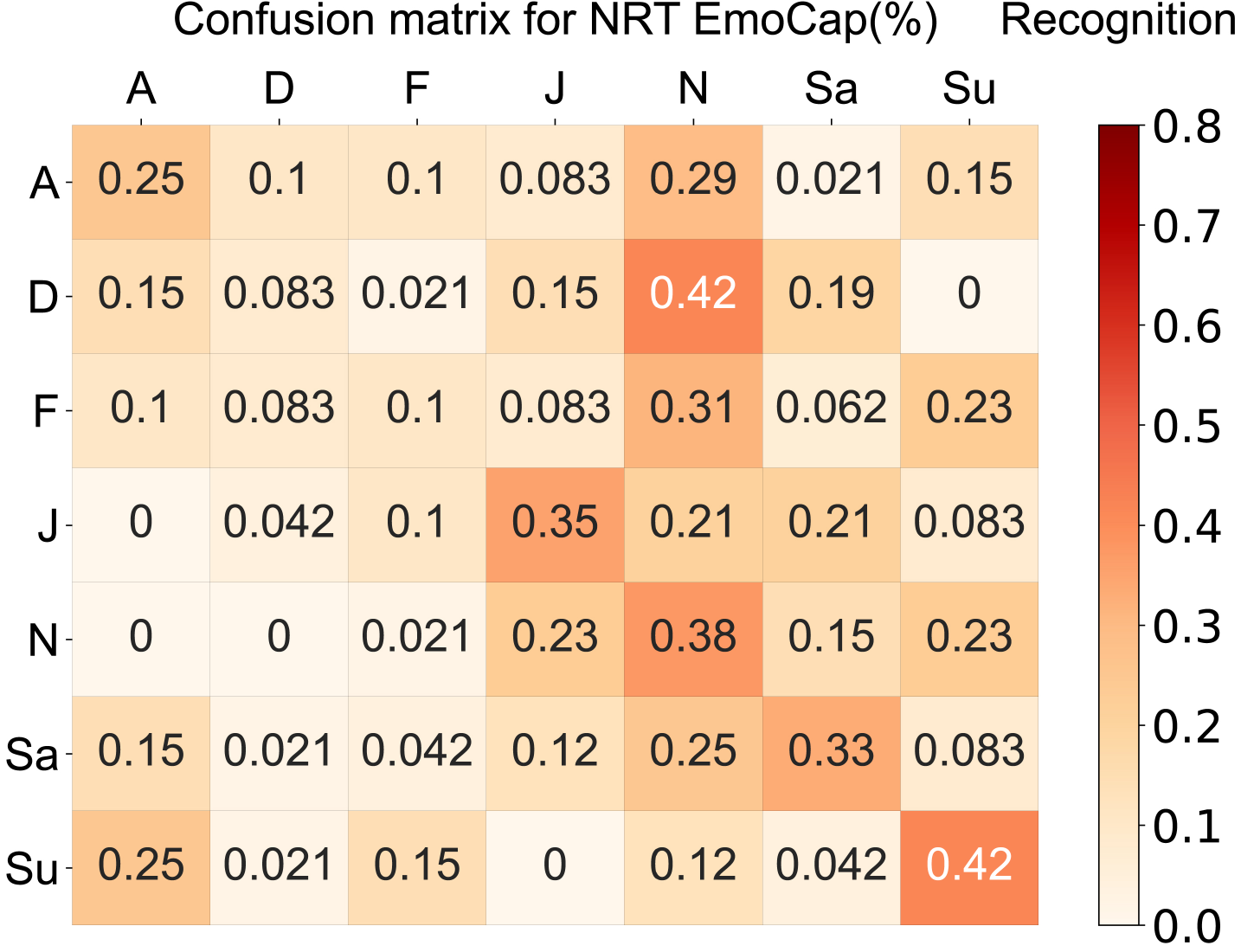}}
    
    \subfigure[RT MienCap for video]{\label{f:realtime_video}\includegraphics[width=0.33\linewidth]{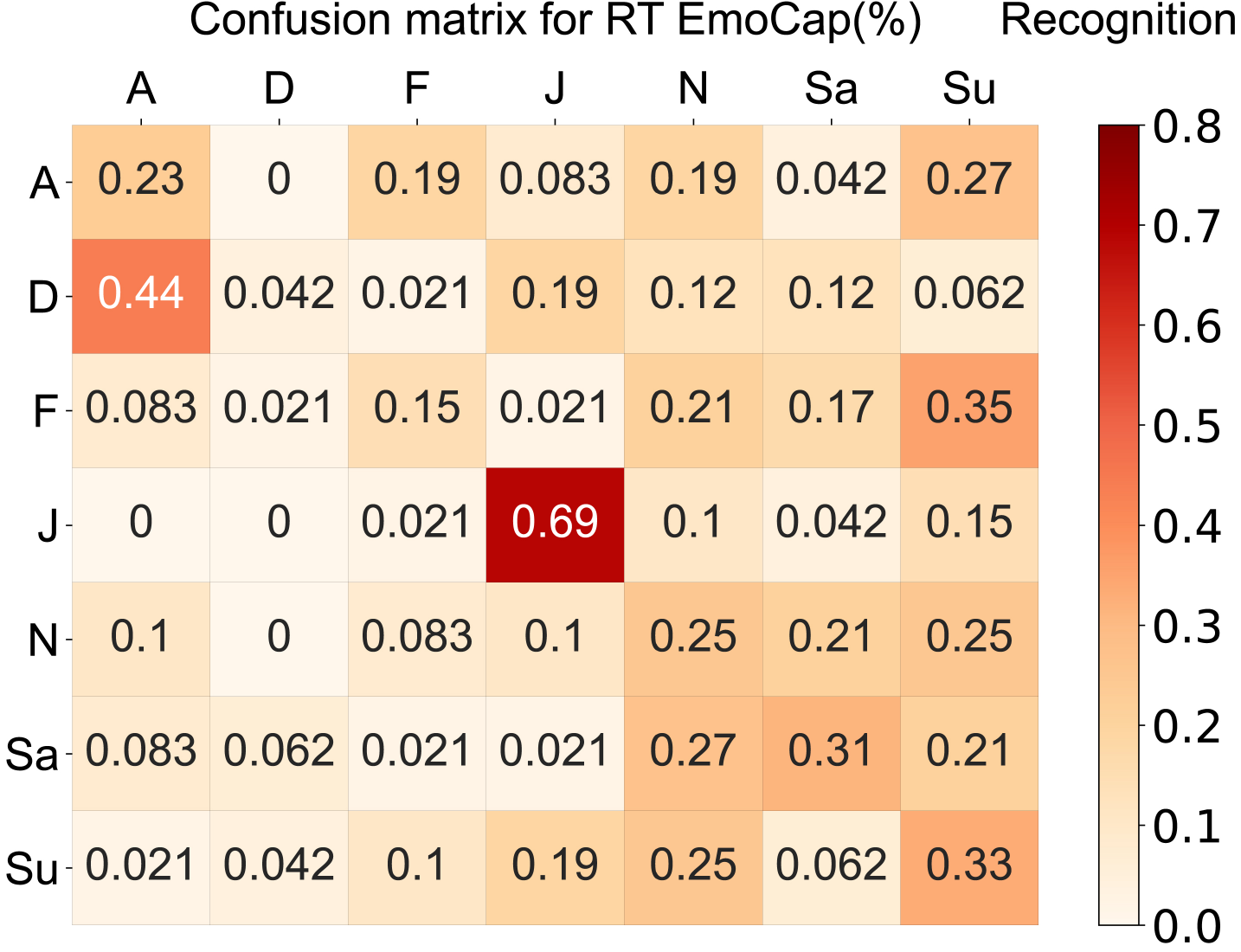}}
    \subfigure[RT MienCap for image]{\label{f:realtime_image}\includegraphics[width=0.33\linewidth]{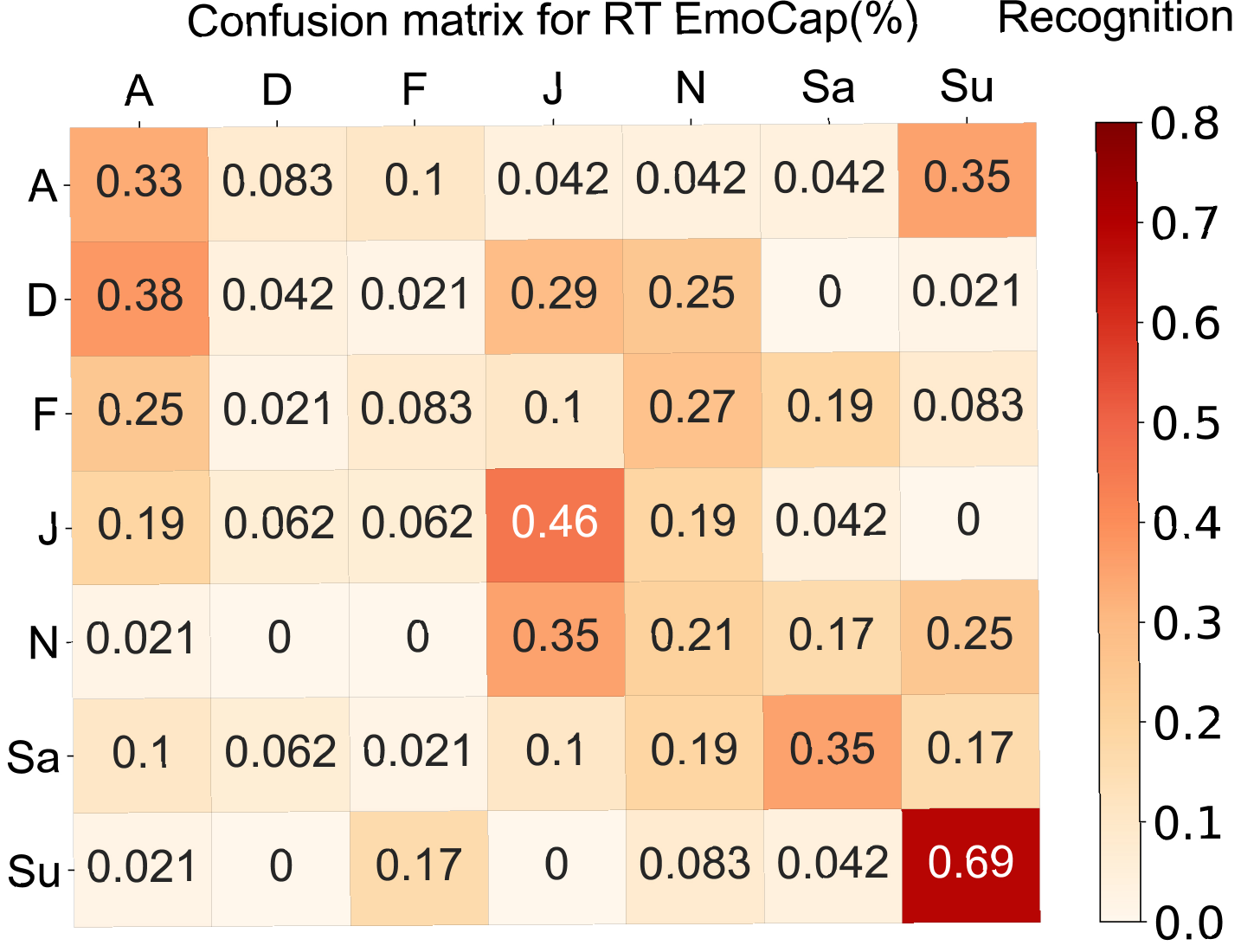}}
  \end{center}
  \caption{Confusion matrix for perceived expression
recognition (\%) for basic expression classes.}
  \label{f:heatMap}
\end{figure*}

\subsection{Scoring \& results}

We applied separate repeated measures Analysis of Variances (ANOVAs) for both video and image, looking at the results on recognition, intensity and attractiveness. Each ANOVA had the within-participants factors character (4), emotion (7), and tracking methods (3). There were no outliers, and the data was normally distributed for each condition as assessed by boxplot and Shapiro–Wilk test ~$(p>0.05)$, respectively. We conducted Mauchly's test to assess sphericity of the data, and whenever it is violated, we report results applied with Greenhouse-Geisser correction and marked with an asterisk ``$^{\ast} $’’. Bonferroni test performed as Post Hoc test for multiple comparisons of means.

\subsubsection{Recognition}
For the recognition of expressions, responses were converted to scores, ``1'' for correct or ``0'' for incorrect, and then averaged over stimuli repetitions. 

\textbf{Videos} Figure ~\ref{f:recognition_video} shows the comparison of average scores obtained for three tracking methods across four characters. The average score over all characters for NRT $(M = .321)$ and RT $(M = .348)$ are significantly higher than the average score for Faceware $(M = .116)$. Firstly, the main effect of tracking method was significant, $ F (2, 22) = 26.531, p < .001$. Bonferroni post-hoc comparisons indicated the mean recognition rates for Faceware is significantly lower than NRT \textcolor{ruisi}{MienCap},~$p < .001$ and RT \textcolor{ruisi}{MienCap},~$p < .001$. However, the mean for NRT \textcolor{ruisi}{MienCap} did not significantly differ from RT \textcolor{ruisi}{MienCap},~$p > .05$. Secondly, tracking methods $\times$ characters interaction, tracking methods $\times$ emotions interaction and tracking methods $\times$ characters $\times$ emotions interaction were not significant, $F(6, 66) = 1.004, p = .43$, $F(3.654, 40.192) = 1.717 , p = .17 ^{\ast} $, $F(6.732, 74.057) = .999, p = .437 ^{\ast} $, respectively.  

\textbf{Images}
Figure ~\ref{f:recognition_image} shows the results for images on the recognition scores. It confirms that the average score over all characters for NRT $(M=.515)$ and RT $(M=.396)$ are significantly higher than the average score for Faceware $(M=.146)$. Firstly, the main effect of tracking method was significant, $ F (2, 22) = 42.094, p < .001$. Bonferroni post-hoc comparisons indicated the mean recognition rates for Faceware is significantly lower than NRT \textcolor{ruisi}{MienCap},~$p < .001$ and RT \textcolor{ruisi}{MienCap},~$p < .001$. However, the mean for NRT \textcolor{ruisi}{MienCap} did not significantly differ from RT \textcolor{ruisi}{MienCap},~$p =.069$. Secondly, tracking methods $\times$ characters interaction, tracking methods $\times$ emotions interaction were not significant, $F(6, 66) = .403, p = .875$, $F(3.217, 35.39) = 2.409, p = .079^{\ast}$, respectively. However, tracking methods $\times$ characters $\times$ emotions interaction was significant, $F(6.793, 74.727) = 2.639, p = .018 ^{\ast} $.

\subsubsection{Intensity}

As expected, intensity ratings for our NRT and RT systems were high, because the facial expressions for stylized characters are generally exaggerated.

\textbf{Videos} Figure ~\ref{f:intensity_video} shows the mean intensity ratings for three tracking methods across four characters. The average score over all characters for NRT $(M = 5.021)$ and RT $(M = 5.048)$ are significantly higher than the average score for Faceware $(M = 4.286)$. Firstly, the main effect of tracking method was significant, $ F (2, 22) = 62.811, p < .001$. Bonferroni post-hoc comparisons indicated the mean intensity ratings for Faceware is significantly lower than NRT \textcolor{ruisi}{MienCap},~$p = .016$ and RT \textcolor{ruisi}{MienCap},~$p = .017$. However, the mean for NRT \textcolor{ruisi}{MienCap} did not significantly differ from RT \textcolor{ruisi}{MienCap},~$p > .05$. Secondly, tracking methods $\times$ characters interaction, tracking methods $\times$ emotions interaction and tracking methods $\times$ characters $\times$ emotions interaction were not significant, $F(2.045, 22.496) = 1.745, p = .197 ^{\ast}$, $F(4.91, 54.009) = .412 , p = .835 ^{\ast} $, $F(7.339, 80.725) = 1.13, p = .353 ^{\ast} $, respectively.  

\textbf{Images} Figure ~\ref{f:intensity_image} shows the results for images on the intensity ratings. The average score over all characters for NRT $(M = 4.765)$ are significantly higher than the average score for Faceware $(M = 4.107)$ and RT $(M = 4.336)$. Firstly, the main effect of tracking method was significant, $ F (2, 22) = 6.259, p = .007$. Bonferroni post-hoc comparisons indicated the mean intensity ratings for RT \textcolor{ruisi}{MienCap} is significantly higher than NRT \textcolor{ruisi}{MienCap},~$p = .026$ and Faceware,~$p = .037$. However, the mean for RT \textcolor{ruisi}{MienCap} did not significantly differ from Faceware,~$p = .83$. Secondly, tracking methods $\times$ characters interaction, tracking methods $\times$ emotions interaction and tracking methods $\times$ characters $\times$ emotions interaction were not significant, $F(2.827, 31.1) = .671, p = .568 ^{\ast}$, $F(4.387, 48.255) = .819 , p = .529 ^{\ast} $, $F(6.855, 75.4) = 1.258, p = .283 ^{\ast} $, respectively.  
\subsubsection{Appeal}
We look at the effect of tracking methods on appeal ratings across all characters. 

\textbf{Videos} Figure ~\ref{f:appeal_video} shows the mean appeal ratings for three tracking methods across four characters. The average score over all characters for NRT $(M = 4.568)$ are significantly higher than the average score for Faceware $(M = 4.158)$. However, the mean for RT \textcolor{ruisi}{MienCap} $(M = 4.524)$ is not significantly different than either these two conditions. Firstly, the main effect of tracking method was significant, $ F (2, 22) = 6.259, p = .007$. Bonferroni post-hoc comparisons indicated the mean appeal ratings for NRT \textcolor{ruisi}{MienCap} is significantly higher than Faceware,~$p = .036$. Secondly, tracking methods $\times$ characters interaction, tracking methods $\times$ emotions interaction and tracking methods $\times$ characters $\times$ emotions interaction were not significant, $F(3.211, 35.317) = .443, p = .737$, $F(4.88, 53.68) = 1.621 , p = .172 ^{\ast} $, $F(7.702, 84.725) = 1.368, p = .224 ^{\ast} $, respectively.  

\textbf{Images} Figure ~\ref{f:appeal_image} shows the results for images on the appeal ratings. The average score over all characters for NRT $(M=4.741)$ and the RT $(M=4.503)$ are significantly higher than the average score for Faceware $(M=4.024)$. Firstly, the main effect of the tracking method was significant, $ F (1.209, 13.304) = 12.859, p = .002$. Bonferroni post-hoc comparisons indicated the mean appeal ratings for Faceware is significantly lower than NRT \textcolor{ruisi}{MienCap},~$p = .005$ and RT \textcolor{ruisi}{MienCap},~$p = .045$. The mean for RT \textcolor{ruisi}{MienCap} is also significantly different from NRT \textcolor{ruisi}{MienCap},~$p =.01$. Secondly, tracking methods $\times$ characters interaction, tracking methods $\times$ emotions interaction, and tracking methods $\times$ characters $\times$ emotions interaction were not significant, $F(2.776, 30.54) = .264, p = .836^{\ast}$, $F(4.62, 50.825) = 1.289, p = .285^{\ast}$, and $F(5.879, 64.673) = .791, p = .578 ^{\ast} $, respectively. 

\section{Discussion}
\subsection{Expression recognition for seven expression classes}
The main effect of emotions was significant, $ F (2.742, 60.329) = 5.531, p = .003$, according to our preliminary data on expression recognition. Thus, we look into participants' rating for seven expression classes. Figure~\ref{f:heatMap} depicts the confusion matrix for each expression class's perceived expression recognition. For a specific row (e.g., anger) in each sub-figure, the columns show the percentage (e.g., averaging nearly over all observed individual anger expressions) of respondents agreeing on the associated expression classes.

\textbf{Human}
We included the human videos and images as the ground truth. The confusion matrix of observed expression recognition rate for humans is shown in Figures 5(a) and 5(b). Surprise and joy are highly accurate, while fear and disgust are extremely difficult for people to recognize and express. This is similar to Aneja et al.'s ~\cite{aneja2018learning} prior result of evaluating human photos as input. We also noticed that character expression identification accuracies are sometimes higher than human, which could be due to the characters' simpler geometry and stylization, which makes the expressions simpler to discern.

\textbf{Faceware}
Figure~\ref{f:blendshape_video} and Figure~\ref{f:blendshape_image} show that the majority of expressions are incorrectly perceived as neutral for Faceware condition. This indicates blendshape based approaches often produce ambiguous expressions, due to the constraints of correspondence mapping.

\textbf{NRT MienCap \& RT MienCap}
Figure~\ref{f:emotion_video} to Figure~\ref{f:recognition_image} show that our NRT and RT system results in more precise expression transfer for the majority of expression classes when compared to Faceware. The most prevalent blunders are mixing up fear and surprise, as well as disgust and fury. Because the confounded statements have similar looking geometric arrangements, these errors are intuitively understandable. Disgust and fear were the least accurate results, as these reactions are difficult to discern in both human and character depictions.

\subsection{Videos \& images}
We took videos from the RAVDESS~\cite{facedataset} stimuli. We note our expression recognition ratings for human is different from the results reported by RAVDESS where their validation tasks were used with North American participants. There is strong evidence for a ‘in-group' advantage in emotion recognition, with accuracy being higher for facial expressions and identified by people from the same cultural group~\cite{elfenbein2002universality}.

\subsection{Animated character emotional audio-visual dataset}
We built the stylized character emotional audio-visual dataset based on RAVDESS~\cite{facedataset} via our real-time MienCap system. This is the first video dataset with animated stylized characters (2 male and 2 female) talking with seven basic emotions. The set consists of synchronized 3D rig parameters, synchronized blendshape weights, synchronized RAVDESS input video and synchronized RAVDESS audio presented in North American English. All available in high-definition formats. Our user study revealed test-retest reliability and high rates of emotional validity. This set may be of interest to a wide range of technologists and researchers.

\subsection{Characters}
We used `Mery’ as the base character, `Bonnie’, `Ray’ \& `Malcolm’ as the secondary character. Our results show the tracking methods $\times$ emotions interaction were not significant in terms of expression recognition, intensity and appeal. This indicates our solutions were effective irrespective of whether that character is a primary or secondary.

Our future work plans to add the concept of a universal base character rig that is powerful enough to create a full range of expressions and can be readily expanded to any new secondary characters.

\section{Conclusion}
In summary, for the first time, we contribute a real time system that captures human facial expressions to drive a stylized character in a perceptually correct and geometrically cohesive fashion. We conducted a user survey to show that our system creates more perceptually accurate expressions than popular commercially accessible software applications, such as, Faceware.

The ease of use of the MienCap system and the expressiveness of our resulting animations can potentially improve the effectiveness of visual storytelling in areas of online marketing, gaming, animated films, and immersive experiences. Our system can also be used in real-time (``live") animation situations, where facial expression is a useful input modality and amateurs can communicate stories with expressive animation by capturing their own performances.

\acknowledgments{
This project has received funding from the European Union’s Horizon 2020 research and innovation programme under grant agreement No. 101017779.}

\bibliographystyle{abbrv-doi}

\bibliography{template}
\end{document}